# Photo-induced Hall Effect in Graphene
## ---Effect of Boundary Types

**Takashi Oka, and Hideo Aoki**

Department of Physics, University of Tokyo, Hongo, Tokyo 113-0033, Japan

oka@cms.phys.s.u-tokyo.ac.jp

**Abstract**. We theoretically predict, with the Keldysh Green's function combined with the Floquet method, that the AC electric field of a circularly polarized light should induce a Hall effect, in the absence of uniform magnetic fields, in graphene with a pair of Dirac dispersions. Although the Hall coefficient is not quantized, the dynamical gap generated by the AC field and the associated Hall effect bear a topological origin which can be traced back to the Dirac cones. We also study the dependence on boundary conditions. The required AC field strength is estimated to be realistic.

### 1. Introduction: a photovoltaic Hall effect

Non-linear phenomena in electronic systems are fascinating since they can lead to transport properties qualitatively distinct from those in equilibrium. In ref. [1], the present authors have predicted a novel mechanism for the Hall effect without magnetic fields. The mechanism --- the "photovoltaic Hall effect" --- takes place in electron systems with Dirac dispersions such as graphene subject to intense, circularly-polarized laser lights. Its physical origin is a charge pumping process accompanying the non-adiabatic evolution of *k*-points in the Brillouin zone induced by the field (Fig.1 (a)). Indeed, when the *k*-point encircles a Dirac point, the wave function acquires a geometric phase known as the Aharonov-Anandan phase --- a non-adiabatic generalization of the Berry phase. This leads, in the Floquet formalism for AC fields, to a gap opening in the Floquet quasi-energy spectrum at the Dirac point which, through the non-equilibrium Kubo formula, contributes to the Hall coefficient.

The Hall coefficient in the presence of the AC electric field background was obtained in [1] for the first time, and is expressed as

$$\sigma_{xy}(A_{ac}) = e^2 \int \frac{dk}{(2\pi)^d} \sum_{\alpha} f_{\alpha}(k) [\nabla \times A_{\alpha}(k)]_z ,$$

where the fictitious gauge field $A_{\alpha}(k)$ arises in the Floquet states and $f_{\alpha}(k)$ the occupation fraction (see [1] for details). Unlike a system in equilibrium where the occupation fraction is universally the Fermi distribution, the occupation fraction in non-equilibrium systems are non-universal but depends on the details of the system, e.g., connection to electrodes, dissipations, etc. However, the term known as the "Berry curvature", $[\nabla \times A_{\alpha}(k)]_z$, exhibits universal features as shown in Fig. 1(b). This causes a finite AC field to produce a photo-induced peak at the Dirac point. In graphene, which has a pair of Dirac points, the photovoltaic Hall effect should take place when (i) the Berry curvature becomes non-zero, and (ii) the occupation fraction for the K and K' points become asymmetric. The typical strength of the AC electric field that is needed is $F=10^7$ V/m for photon energy of $\Omega =O(0.1)$ eV.

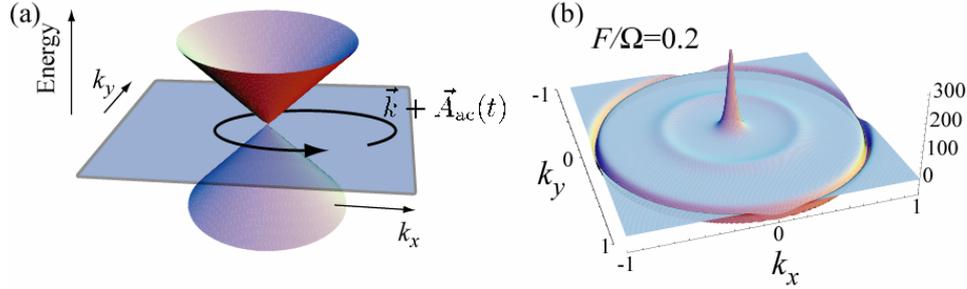

Fig. 1: (color online) (a) A trajectory of $k+A_{ac}(t)$ around the Dirac point driven by a circularly polarized light. $A_{ac}(t)=F/\Omega$ (cos $\Omega t$, sin $\Omega t$) is the AC electric field where $F$ is the strength and $\Omega$ the photon energy. (b) The photo-induced Berry curvature in $k$-space [1].

## 2. Dynamical gap in graphene and its dependence on the boundary

An important effect of the AC field is the Wannier-Stark localization: The field opens dynamical gaps in the spectrum [1, 2]. In the case of linearly polarized lights, the relevant gaps appear at $\omega = \Omega/2$ and $\omega = -\Omega/2$, whereas in circularly polarized lights an extra gap opens at the Dirac point $\omega=0$ as well. Since the gap at the Dirac point is deeply connected to the Berry curvature, the photovoltaic Hall effect only takes place in circularly (or elliptically) polarized lights.

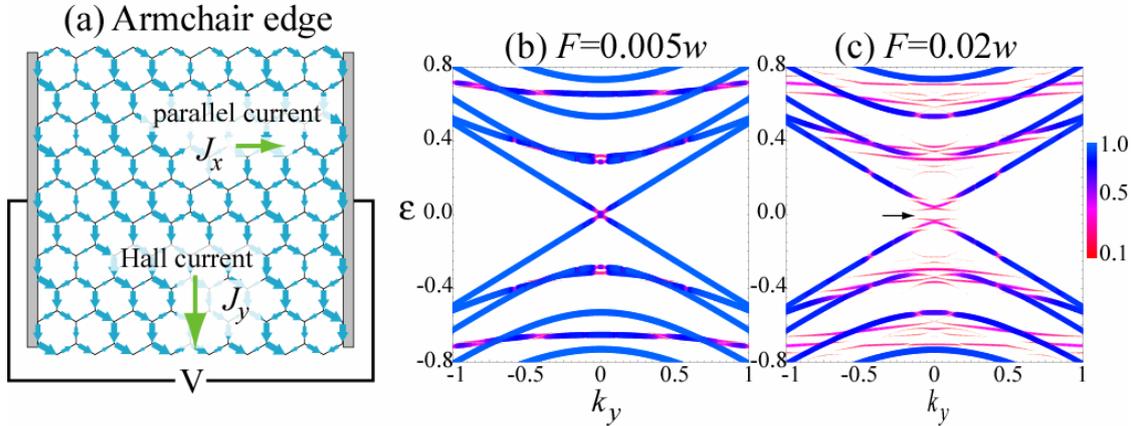

Fig. 2: (color online) (a) DC current distribution in graphene subject to a circularly polarized light of intensity $F=0.025w$ and photon energy $\Omega =0.3w$ ($w$: the transfer integral of graphene) with electrodes attached to armchair edges with a bias $V=0.005w$. (b), (c) Floquet quasi-energy dispersion for $F=0.005w$ (b) and $0.02w$ (c) with $\Omega =0.3w$. The arrow indicates the dynamical gap at the Dirac point.

We have numerically calculated the transport properties of a graphene ribbon attached to two electrodes with bias $V$ subject to strong ac fields by combining the Keldysh Green's function method with the Floquet method [1]. In Fig. 1, we plot the results for electrodes attached to armchair edges. In finite ac fields, we find that DC current is no longer parallel to the bias but acquires a perpendicular component, which confirms that photovoltaic Hall effect indeed takes place. The Floquet energy dispersions are shown in Fig. 1 (b), (c), where we can see that a gap clearly opens at the Dirac point.

We can actually confirm that the Hall coefficient and the dynamical gap are intimately related: Numerical results suggest that the Hall coefficient is proportional to the gap.

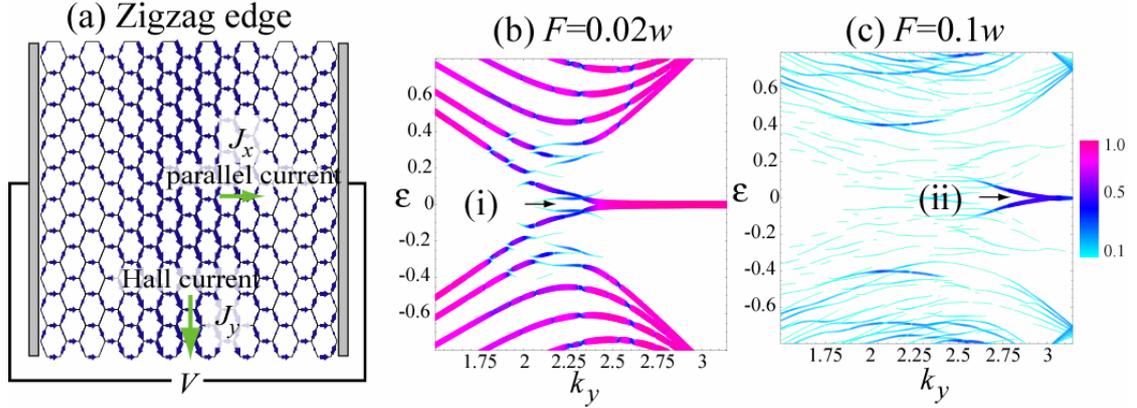

Fig. 3: (color online) (a) DC current distribution in graphene with electrodes attached to zigzag edges for $F=0.02w$, $\Omega=0.3w$ and $V=0.3w$. (b), (c) Floquet quasi-energy dispersions in the gap-closing region of the BZ around $k_y=2/3\pi$ for $F=0.02w$ (b) and $F=0.1w$ (c), with photon energy $\Omega=0.3w$. The arrows (i) (ii) indicate the dynamical gap opening for the flat portion in the dispersion.

In Fig. 3, we plot the result when the electrodes are attached to zigzag edges. The electric-field dependence of the Hall current is qualitatively the same with the armchair case. A distinct feature, however, lies in the quasi-energy spectrum. For zigzag edges a flat band exists in the spectrum which originates from the edge state. As indicated by arrows (i), (ii) in Fig. 3 (b)(c), the gap opening makes the flat portion of the band shorter as the field strength increase and disappears at $F > \pi\Omega/3$, i.e., when $k_y + A_{ac}(t)_y$ at the Dirac point ($k_y=2/3\pi$) touches the Brillouin zone boundary. However, since the photovoltaic Hall effect is a bulk effect, i.e., the current flows in the center of the system as shown in Fig. 3 (a), the contribution from the edge state to the photovoltaic Hall current is insignificant.

## 3. Conclusion

In the present article, we have extended our analysis on the photovoltaic Hall effect in graphene, first presented in ref. [1], by studying the transport properties as well as the Floquet quasi-energy spectrum in two edge types, namely the armchair and the zigzag edges. In both edge types, photovoltaic Hall effect is observed numerically when the applied circularly polarized AC field is strong enough, whereas the spectra exhibit distinct features. In the zigzag edge case, the flat band of the edge state persists until the strength of the ac field exceeds a critical value at which the *k*-point around the Dirac point reaches the Brillouin zone boundary. Given the required AC field strength estimated here is realistic, we emphasize that the photovoltaic Hall effect predicted here can be experimentally verified.

**References**
[1] T. Oka, and H. Aoki, Phys. Rev. B **79**, 081406 (R) (2009).
[2] S. V. Syzranov, M. V. Fistul, K. B. Efetov, Phys. Rev. **B 78**, 045407 (2008)